\documentclass[11pt,a4paper]{article}

\pdfoutput=1

\usepackage{amsmath,amssymb}
\usepackage{natbib}
\usepackage[colorlinks=true]{hyperref}
\usepackage{graphicx}
\usepackage[position=top]{subfig}
\usepackage[a4paper]{geometry}

\begin{document}

\title{Blocking Collapsed Gibbs Sampler for Latent Dirichlet Allocation Models}

\author{
    X. Zhang\thanks{School of Mathematics and Statistics, University of New South Wales, Sydney 2052 Australia.}
    \thanks{Communicating Author: {\tt x.zhn421@yahoo.com}}
    \and
    S. A. Sisson\footnotemark[1]
    }

\date{}

\maketitle

\begin{abstract}
The latent Dirichlet allocation (LDA) model is a widely-used latent variable
model in machine learning for text analysis. Inference for this model
typically involves a single-site collapsed Gibbs sampling step for
latent variables associated with observations. The efficiency of the
sampling is critical to the success of the model in practical large
scale applications. In this article, we introduce a blocking scheme
to the collapsed Gibbs sampler for the LDA model which can, with
a theoretical guarantee, improve chain mixing efficiency. We develop two procedures, 
an $O(K)$-step backward simulation and an $O(\log{K})$-step nested 
simulation, to directly sample the latent variables within each block. 
We demonstrate that the blocking scheme achieves substantial improvements in chain 
mixing compared to the state of the art single-site collapsed Gibbs sampler. 
We also show that when the number of topics is over hundreds, the nested-simulation  
blocking scheme can achieve a significant reduction in computation 
time compared to the single-site sampler.
\end{abstract}

\section{Introduction}

Gibbs sampling is an iterative scheme to generate random samples from
a posterior distribution, which has underpinned
many important applications in Bayesian statistics and machine learning \citep{Andrieu:2003}.
It is applicable when the joint distribution is difficult to sample
from directly, but the distribution of each variable conditional on
the rest, is known and is easy to simulate from. The Gibbs sampler
often works in a single-site update \citep{Geman:1984} or data
augmentation \citep{Tanner:1987} manner. Multiple
sampling techniques may be applicable, such as collapsing and blocking,
which are able to improve chain mixing \citep{Liu:1994a}. In this
article, we propose a blocking scheme to improve the efficiency of
the collapsed Gibbs sampler for the latent Dirichlet allocation (LDA)
model, which is popular for topic modelling. We demonstrate that the
proposed sampler achieves substantial improvements 
compared to the state of the art single-site collapsed Gibbs sampler
\citep{Griffiths:2004}.

The LDA model is a Bayesian hierarchical mixture model, which posits
a fixed number of topics (mixture components) for a collection of
documents, known as a corpus. It assumes that each document in the
corpus reflects a combination of those topics. The model is then used to
extract those unknown topics from a given corpus. In the model, each
topic is characterised by a distinct multinomial topic-specific distribution,
over a typically large vocabulary, while each document is modelled
by a multinomial document-specific distribution over all topics. Thus,
the distribution of the words from one document is a mixture of multinomial distributions
over the vocabulary. The sharing of mixture components and the varying
of mixture coefficients among documents reveals the similarity or
dissimilarity of the underlying patterns of their words. 

\citet{Blei:2003} first proposed the LDA model to find the underlying
patterns of words from corpora. Finding these patterns allows for
effective corpus exploration, document classification, and information
retrieval. It has multiple applications in areas such as text processing
\citep{Blei:2003,Griffiths:2004} and computer vision \citep{FeiFei:2005}.
In practical text analysis applications, LDA models have previously
been fitted to corpera containing more than tens of thousands of
documents, for vocabularies of over tens of thousands of unique terms
and for hundreds of topics. This leads to models with many millions
of parameters, which is a considerable challenge for Bayesian inference. 

Many methods have been developed for inference and learning, such
as variational methods \citep{Minka:2002,Blei:2003,Teh:2006b} and
the collapsed Gibbs sampling method \citep{Griffiths:2004}. The collapsed
Gibbs sampler generates word-topic allocations for all words in the
corpus. Their conditional sampling distributions are derived by integrating
out the multinomial parameters of the document-specific distributions
as well as those of topic-specific ones. The topic allocation of each
word is updated sequentially w.r.t. a discrete distribution
over all topics, i.e. the sampler performs single-site updates.
This collapsed Gibbs sampler has been shown to achieve better results
faster than variational methods on small to medium corpora 
\citep{Griffiths:2004,Teh:2006b,Asuncion:2009}. 
Beyond Gibbs sampling, some researchers \citep{Welling:2011,Ahn:2012,Patterson:2013} have
proposed using Langevin Monte Carlo methods combined with stochastic
gradient techniques for posterior inference. These approaches can produce
faster sampler updates, as they are only constructed
from a subset of observations in each iteration. However, they have worse performance
in chain mixing.

While the collapsed Gibbs sampler employs Rao\textendash Blackwellization
\citep{Casella:1996} to avoid explicitly sampling some parameters,
it can however exhibit slow mixing because it only updates one hidden
state assignment at a time \citep{Celeux:2000}. As such, its performance
deteriorates quickly when working with large datasets, which are typical
in text analysis. 

Various attempts have been made to scale up the collapsed Gibbs sampler
to analyse increasingly large scale document corpora. Some
researchers have proposed developing sampling strategies that can
mimic the collapsed Gibbs dynamic, under distributed or online mini-batch
settings \citep{Smyth:2008,Newman:2009,Canini:2009}. These approaches can provide
substantial memory and time savings, but they are not guaranteed to
sample from the true posterior distribution. 

In collapsed Gibbs sampling, the cost of evaluating and simulating
from discrete distributions, which have the same dimension as the
number of topics, consumes a major part of the overall computation
time. Several authors \citep{Porteous:2008b,Yao:2009,Li:2014,Yuan:2015}
have investigated different approaches to reduce such computational
complexity. Their methods either exploit the sparsity of observations,
and/or use multiple cheap independent Metropolis proposals \citep{Andrieu:2003}
instead of the expensive full conditional sampling distributions. Though these
approaches have provided some improvements to the computation time,
their sampling efficiency is ultimately hindered by the mixing rate
of the single-site collapsed Gibbs sampler.

In conclusion, most of the existing work on Markov chain Monte Carlo (MCMC) methods for the LDA
model attempts to achieve faster operations on computation by
retaining or sacrificing the chain mixing efficiency of the collapsed
Gibbs sampler. In this article, we propose a non-trivial blocking
scheme for the collapsed Gibbs sampler, which is theoretically
guaranteed to accelerate chain mixing \citep{Liu:1994a}. We first
provide the background of the model and discuss existing sampling approaches
in Section~\ref{sec:bg}. In Section~\ref{sec:blk}, we introduce the
blocking scheme, from which the full conditional distributions of the
blocked latent variables can be directly simulated. We develop   
 $O(K)$-step backward simulation and $O(\log{K})$-step nested 
simulation schemes to achieve this.
We examine the
performance of the proposed blocking scheme for one simulated 
and two real world datasets in Section~\ref{sec:ex}, and demonstrate
that the proposed sampler can achieve substantial improvements in
chain mixing, compared to the state of the art single-site collapsed
Gibbs sampler. Regardless of its quadratic  
computational complexity in evaluating sampling densities, the nested-simulation 
blocking scheme can also achieve a reduction in computation time per iteration when 
there are more than a few hundred topics.
In Section~\ref{sec:disc}, we discuss some
possible future research directions.

\section{Background}
\label{sec:bg}

We first provide a brief review of the LDA model and its associated
Gibbs sampling approaches.

\subsection{Model}

The LDA model summarises a document collection by multiple topics,
where each topic may potentially span multiple documents. A standard
assumption is that the data are exchangeable, i.e.~the order
of documents in a collection does not matter, and that the order of
words in a document does not matter. 

Let $w_{dn}\in\{1,\ldots,V\}$ be the word at 
position $n$ in document $d$, with its value indicating
a word from a vocabulary of size $V$. Document $d$
of length $N_{d}$ is then constructed as $w_{d}=(w_{d1},\ldots,w_{dN_{d}})$.
Given $K$ topics, the topic-specific distribution for topic $k$
is a $V$-dimensional multinomial distribution with parameter vector
$\phi_{k}=(\phi_{k1},\ldots,\phi_{kV})$, $0\leq\phi_{kv}\leq1$ for
all $v$ and $\sum_{v}\phi_{kv}=1$. That is, for topic $k$,
the probability of observing $w_{dn}=v$ is $\phi_{kv}$.

Similarly, let $z_{dn}\in\{1,\ldots,K\}$ be the latent variable for each word
$w_{dn}$. Its value denotes the topic to which the associated word $w_{dn}$
belongs. The value of $z_{dn}$ follows the document-specific distribution
for document $d$, which is a multinomial distribution
with parameter vector $\theta_{d}=(\theta_{d1},\ldots,\theta_{dK})$,
$0\leq\theta_{dk}\leq1$ for all $k$ and $\sum_{k}\theta_{dk}=1$.
In document $d$, the probability that $z_{dn}=k$,
i.e. the probability that $w_{dn}$ is associated with topic $k$,
is $\theta_{dk}$.

Let $\phi_{1:K}=(\phi_{1},\ldots,\phi_{K})$ be the multinomial parameters
of the topic-specific distributions for all $K$ topics and $z_{d}=(z_{d1},\ldots,z_{dN_{d}})$
be the labels for all words in document $d$. In
the LDA model, the likelihood function for $w_{d}$ is a mixture of
$K$ components, which are the topic-specific distributions, with
mixture coefficients $\theta_{d}$. This mixture structure formation
leads to a latent variable model, with the complete likelihood function
\begin{equation}
\label{eq:lik-complete-word}
p(w_{d},z_{d}|\theta_{d},\phi_{1:K})=
    \prod_{n}\theta_{d,z_{dn}}\phi_{z_{dn},w_{dn}}.
\end{equation}

If there are $D$ documents, then their words are $w_{1:D}=(w_{1},\ldots,w_{D})$
and their associated topics are $z_{1:D}=(z_{1},\ldots,z_{D})$. Let
$\theta_{1:D}=(\theta_{1},\ldots,\theta_{D})$ be the multinomial
parameters of the document-specific distributions for all  $D$
documents. To proceed with Bayesian inference, we specify the Dirichlet
distribution $\pi(\theta_{d}|\alpha_{1:K})$, with $\alpha_{1:K}=(\alpha_{1},\ldots,\alpha_{K}),\,\alpha_{k}>0$
for all $k$, as the prior for $\theta_{d}$, and the symmetric Dirichlet
distribution $\pi(\phi_{k}|\beta),\,\beta>0$ as the prior for
$\phi_{k}$. The posterior distribution can then be obtain from (\ref{eq:lik-complete-word}) as
\begin{equation}
\label{eq:posterior-joint-word}
\pi(z_{1:D},\theta_{1:D},\phi_{1:K}| w_{1:D},\alpha_{1:K},\beta)\propto
    \prod_{d}\prod_{k}\theta_{dk}^{S_{d\cdot k}+\alpha_{k}-1}\times
    \prod_{k}\prod_{v}\phi_{kv}^{S_{\cdot vk}+\beta-1},
\end{equation}
where $S_{d\cdot k}$ is the number of words in 
document $d$ associated with topic $k$ and $S_{\cdot vk}$ is the number
of words in all documents taking the value $v$ and being associated
with topic $k$. Both $S_{d\cdot k}$ and $S_{\cdot vk}$ are functions
of $w_{1:D}$ and $z_{1:D}$.

\subsection{Gibbs sampling }

Due to the latent variable structure (\ref{eq:posterior-joint-word}),
a data augmentation scheme, under which the sampler targets both latent
variables $z_{1:D}$ and parameters $(\theta_{1:D},\phi_{1:K})$,
can be naturally devised, by alternating simulation between Dirichlet
densities and discrete densities as follows:
\begin{eqnarray}
\label{eq:sampler-augment}
\theta_{d}| z_{d}\sim\prod_{k}\theta_{dk}^{S_{d\cdot k}+\alpha_{k}-1}\text{ for all }d 
    & \text{and} & \phi_{k}| z_{1:D}\sim\prod_{v}\phi_{kv}^{S_{\cdot vk}+\beta-1}\text{ for all }k;\\
z_{dn}=k|\theta_{d},\phi_{1:K}\sim\frac{\theta_{dk}\phi_{k,w_{dn}}}{\sum_{l}\theta_{dl}\phi_{l,w_{dn}}} 
    &  & \text{for all }(d,n).\nonumber 
\end{eqnarray}

As an alternative, \citet{Griffiths:2004} proposed to use a collapsed
Gibbs sampling scheme, in which the sampler explores the marginal
posterior distribution of the latent variables $z_{1:D}$, given by
\begin{equation}
\label{eq:posterior-collapsed-word}
p(z_{1:D}| w_{1:D},\alpha_{1:K},\beta)\propto
    \frac{\prod_{d}\prod_{k}\Gamma(S_{d\cdot k}+\alpha_{k})\times\prod_{k}\prod_{v}\Gamma(S_{\cdot vk}+\beta)}
    {\prod_{k}\Gamma(S_{\cdot\cdot k}+V\beta)},
\end{equation}
where $S_{\cdot\cdot k}=\sum_{v}S_{\cdot vk}$ is the number of words
associated with topic $k$ in all documents. This statistic is also a function
of $w_{1:D}$ and $z_{1:D}$.

We denote $z_{1:D}^{(-dn)}$ to mean $z_{1:D}$, but excluding the
single element $z_{dn}$, with an analogous definition for $w_{1:D}^{(-dn)}$.
The single-site collapsed Gibbs sampling approach of \citet{Griffiths:2004}
updates each $z_{dn}$ sequentially, conditional on the remaining
latent variables, from a $K$-dimensional discrete distribution 
\begin{equation}
\label{eq:sampler-collapsed}
p(z_{dn}=k| z_{1:D}^{(-dn)},w_{1:D},\alpha_{1:K},\beta)\propto
    \frac{(S_{d\cdot k}^{(-dn)}+\alpha_{k})(S_{\cdot vk}^{(-dn)}+\beta)}{S_{\cdot\cdot k}^{(-dn)}+V\beta},
\end{equation}
where $S_{d\cdot k}^{(-dn)}$, $S_{\cdot vk}^{(-dn)}$ and $S_{\cdot\cdot k}^{(-dn)}$,
respectively, denote the values of $S_{d\cdot k}$, $S_{\cdot vk}$
and $S_{\cdot\cdot k}$ constructed from $w_{1:D}^{(-dn)}$ and $z_{1:D}^{(-dn)}$.
\citet{Newman:2009} show empirically that this collapsing scheme
is more efficient than the data augmentation sampler (\ref{eq:sampler-augment}) 
in achieving better predictive performance.

Although this single-site sampler is straightforward and easily to
implement, it can, however, be slow to converge and mix poorly, especially
for models with mixture structures. \citet{Celeux:2000} attributed this mixing problem
to the incremental nature of the single-site Gibbs sampler, which is unable to
simultaneously move a group of variables to a different mixture component.
A sampling scheme which allows a group of latent variables to be updated
simultaneously may remedy this problem, as \citet{Liu:1994a} have
proven that grouping dependent variables can improve 
chain mixing efficiency. Therefore, a blocking scheme within the collapsed
Gibbs sampler should potentially provide a considerable performance
boost.

\section{Blocking}
\label{sec:blk}

In this section, we propose a blocking scheme for the existing collapsed
(\ref{eq:sampler-collapsed}) for the LDA model, which can
improve chain mixing with a theoretical guarantee. We first construct
the sufficient statistic w.r.t. $\theta_{1:D}$ and $\phi_{1:K}$, which naturally
leads to a blocking scheme. We then develop a backward simulation and 
a nested simulation for exact sampling from the full conditional distributions
of the blocked latent variables.

\subsection{Sufficient statistic}

For document $d$, we define $n_{dvk}=\sum_{n}\delta_{(v,k)}(w_{dn},z_{dn})$
to be a statistic of $(w_{d},z_{d})$, where $\delta_{(v,k)}(w_{dn},z_{dn})=1$
if and only if $(w_{dn},z_{dn})=(v,k)$. Hence, $n_{dvk}$ enumerates
the number of times word $v$ is associated with topic $k$ in document $d$.
In this way, we can summarise $(w_{d},z_{d})$ by a $V\times K$
matrix $\mathbb{N}_{d}$ with entries $n_{dvk}$.

In the following, we show that $\mathbb{N}_{d}$ is sufficient
for $(\theta_d,\phi_{1:K})$ in the complete likelihood function
(\ref{eq:lik-complete-word}). Let $n_{d\cdot k}=\sum_{v}n_{dvk}$
be the number of $z_{dn}$ being equal to $k$, and $n_{dv\cdot}=\sum_{k}n_{dvk}$
be the number of times word $v$ appears in document $d$.
Define $C_{dv}=\sum_{n}\sum_{k}\delta_{(v,k)}(w_{dn},z_{dn})$ to
be the total count of word $v$ in document $d$.
Grouping those terms $\theta_{d,z_{dn}}\phi_{z_{dn},w_{dn}}$ for which
$(w_{dn},z_{dn})$ takes the same values $(v,k)$, we can rewrite
the complete likelihood function (\ref{eq:lik-complete-word}) as
\begin{equation}
\label{eq:lik-table-complete}
p(w_{d},\mathbb{N}_{d}|\theta_{d},\phi_{1:K})=
    \prod_{v}\left\{ \frac{C_{dv}!}{\prod_{k}(n_{dvk}!)}\times
    \prod_{k}(\theta_{dk}\phi_{kv})^{n_{dvk}}\right\},
\end{equation}
where $\prod_{v}\frac{C_{dv}!}{\prod_{k}(n_{dvk}!)}$ is the total number of 
equivalent realisations of $z_d$, i.e. those realisations of $z_d$ leading to 
the same $\mathbb{N}_{d}$ given $w_d$.
Note that (\ref{eq:lik-table-complete}) is equal to (\ref{eq:lik-complete-word})
up to a multiplicative constant. Therefore, 
$\mathbb{N}_{d}$ is sufficient for $\theta_d$ and $\phi_{1:K}$.

For all $D$ documents, the corresponding 3-dimensional matrix of
sufficient statistic is $\mathbb{N}_{1:D}=(\mathbb{N}_{1},\ldots,\mathbb{N}_{D})$.
Given the Dirichlet priors for each $\theta_{d}$ and $\phi_{k}$,
the collapsed posterior (\ref{eq:posterior-collapsed-word}) can be derived by integrating
out $\theta_{1:D}$ and $\phi_{1:K}$, which gives
\begin{equation}
\label{eq:posterio-collapsed-table}
    p(\mathbb{N}_{1:D}| w_{1:D},\alpha_{1:K},\beta)\propto
    \frac{\prod_{d}\prod_{k}\Gamma(n_{d\cdot k}+\alpha_{k})\times\prod_{k}\prod_{v}\Gamma(n_{\cdot vk}+\beta)}
    {\prod_{d}\prod_{v}\prod_{k}(n_{dvk}!)\times\prod_{k}\Gamma(n_{\cdot\cdot k}+V\beta)}.
\end{equation}
Due to the sufficiency, all equivalent realisations of $z_{d}$ given $w_{d}$
are uniformly distributed. 
In particular, we consider $B_{dv}=\{z_{dn};\,w_{dn}=v\}$, which is the group of 
$z_{dn}$ in document $d$ with their associated $w_{dn}$ taking word $v$.
Given $(n_{dvk})_{1:K}=(n_{dv1},n_{dv2},\ldots,n_{dvK})$, all the equivalent
realisations of $B_{dv}$ are uniformly distributed.
Therefore, the blocked sampling scheme can be built
upon the above posterior distribution (\ref{eq:posterio-collapsed-table}) 
by sequentially sampling the
blocks $B_{dv}$ via sampling $(n_{dvk})_{1:K}$ for all $d,\,v$.

\subsection{Blocking scheme}

The blocking scheme we consider is to sample all latent variables
$z_{dn}$ in the group $B_{dv}$ simultaneously, conditional on the
rest. To do this, their associated sufficient statistics $(n_{dvk})_{1:K}$
can first be simulated from the full conditional distribution,
\begin{equation}
\label{eq:bcgs-table}
p((n_{dvk})_{1:K}|\text{rest})\propto
    \prod_{k}\frac{(n_{d\cdot k}^{[-dvk]}+\alpha_{k})^{(n_{dvk})}\times(n_{\cdot vk}^{[-dvk]}+\beta)^{(n_{dvk})}}
    {n_{dvk}!\times(n_{\cdot\cdot k}^{[-dvk]}+V\beta)^{(n_{dvk})}},
\end{equation}
where $n_{d\cdot k}^{[-dvk]}=n_{d\cdot k}-n_{dvk}$, $n_{\cdot vk}^{[-dvk]}=n_{\cdot vk}-n_{dvk}$,
$n_{\cdot\cdot k}^{[-dvk]}=n_{\cdot\cdot k}-n_{dvk}$ and 
$x^{(n)}=\Gamma(x+n)/\Gamma(x)=x(x+1)\cdots(x+n-1)$. Next, all
$z_{dn}\in B_{dv}$ are updated jointly, by uniformly choosing from
all possible topic allocations resulting in the same $(n_{dvk})_{1:K}$.
In practice, this last step can be skipped because knowing $(n_{dvk})_{1:K}$
is enough to proceed the subsequent computation. 
If the block contains more than one variable ($C_{dv} > 1$), this blocking scheme 
is theoretically guaranteed to accelerate chain mixing efficiency \citep{Liu:1994a},
as all $z_{dn}\in B_{dv}$ are dependent in the collapsed 
posterior distribution~(\ref{eq:posterior-collapsed-word}).

Direct simulation from (\ref{eq:bcgs-table}) requires
evaluation of its normalising constant. As
this unnormalised density function is the product of $K$ functions
of each $n_{dvk}$, this structure allows for a sum-of-product algorithm \citep{Bishop:2006} for
exact computation of this collection of normalising constants, and 
backward (Section~\ref{sec:blk-back}) or nested (Section~\ref{sec:blk-nested}) simulations 
for direct sampling from the distribution (\ref{eq:bcgs-table}). 

To simplify notation, we rewrite the unnormalised density function
(\ref{eq:bcgs-table}) as $\prod_{k}q_{k}(n_{dvk})$ with 
\[
q_{k}(n_{dvk}) = 
    \frac{(n_{d\cdot k}^{[-dvk]}+\alpha_{k})^{(n_{dvk})}\times(n_{\cdot vk}^{[-dvk]}+\beta)^{(n_{dvk})}}
    {n_{dvk}!\times(n_{\cdot\cdot k}^{[-dvk]}+V\beta)^{(n_{dvk})}},
\]
for $k=1,\ldots,K$. If we let 
\begin{equation}
\label{eq:normalising-constant}
h_{k_0:k_1}(c)={\displaystyle \sum_{n_{dvk_0}+\cdots+n_{dvk_1}=c}}q_{k_0}(n_{dvk_0})\times
\cdots\times q_{k_1}(n_{dvk_1}),
\end{equation}
where $k_0 \leq k_1$, then $h_{1:K}(C_{dv})$ is the normalising constant of $\prod_{k}q_{k}(n_{dvk})$,
so that the full conditional distribution of $(n_{dvk})_{1:K}$ is
\[
p((n_{dvk})_{1:K}|\text{rest})=\frac{\prod_{k}q_{k}(n_{dvk})}{h_{1:K}(C_{dv})}.
\]

\subsubsection{Backward simulation}
\label{sec:blk-back}
Backward simulation works in a sequential manner. It first samples
$n_{dvK}$ from its marginal distribution. Then backwards from $k=K-1,\ldots,1$, it 
samples $n_{dvk}$ from its conditional distribution given
$n_{dvK},n_{d,v,K-1},\ldots,$ $n_{d,v,k+1}$. Provided values of the
normalising constant $h_{1:k}(c)$ for any $k\leq K$ and $c\leq C_{dv}$
are available, such a sampling procedure can be naturally devised
due to the factored structure of (\ref{eq:bcgs-table}). 

First, the number of words in topic $K$ can be directly simulated from the discrete
marginal distribution 
\[
p(n_{dvK}=n|\text{rest})=\frac{q_{K}(n)\times h_{1:(K-1)}(C_{dv}-n)}{h_{1:K}(C_{dv})},
\]
for $n=0,1,\ldots,C_{dv}$. 
Then progressing
backwards, the number of words for topic $k=K-1,\ldots,3$, given those
previously simulated for topics $k+1,\ldots,K$, can be simulated from the distribution
\begin{equation}
\label{eq:sampler-backward}
p(n_{dvk}=n| n_{d,v,k+1},\ldots,n_{dvK},\text{rest})=
    \frac{q_{k}(n)h_{1:(k-1)}(C_{dv}-\sum_{l=k+1}^{K}n_{dvl}-n)}{h_{1:k}(C_{dv}-\sum_{l=k+1}^{K}n_{dvl})}.
\end{equation}
for $n=0,1,\ldots,C_{dv}-\sum_{l=k+1}^{K}n_{dvl}$. 
For the final stage $k=2$,  the configuration for
the first two topics, $(n_{dv1},n_{dv2})$ can be  
simultaneously sampled from the joint distribution
\[
p(n_{dv1}=n_{1},n_{dv2}=n_{2}| n_{dv3},\ldots,n_{dvK},\text{rest})=
    \frac{q_{2}(n_{2})q_{1}(n_{1})}{h_{1:2}(C_{dv}-\sum_{k=3}^{K}n_{dvk})}.
\]
It is trivial to see that the product of these conditional densities
leads to the target density (\ref{eq:bcgs-table}).

To enable backward simulation, we need to be able to compute the normalising constants
$h_{1:k}(c)$ for any $k\leq K$ and $c\leq C_{dv}$. Due to its factored
structure (\ref{eq:normalising-constant}),  a forward
summation approach can be used to recursively obtain each value of $h_{1:k}(c)$.

It is trivial that $h_{1:1}(c)=q_{1}(c)$
for any $c$.
The constants $h_{1:k}(c)$ with $k\geq2$ can be sequentially computed
through the forward recursive equation 
\[
h_{1:k}(c)=\sum_{n=0}^{c}q_{k}(n)\times h_{1:(k-1)}(c-n),
\]
as each constituent term $h_{1:(k-1)}(0)$, $h_{1:(k-1)}(1)$,
$\ldots,$ $h_{1:(k-1)}(c)$ will have been previously calculated.

\subsubsection{Nested simulation}
\label{sec:blk-nested}
Backward simulation costs $O(K)$ steps of discrete sampling. 
As $K$ is large in practice, the blocked sampling 
could be painfully slow such that its gain in chain mixing is worthless. 
Therefore, we propose a nested simulation scheme which takes at most $O(C_{dv}\log{K})$ steps.

In nested simulation, a binary tree is used to represent a nested partition structure of all 
$K$ topics. The root takes all topics, 
with its left-child node taking those for topic 1 to topic $K_{1/2}=[(K+1)/2]$ 
(where $[x]$ denotes the integer part of $x$), while its right-child node takes the rest of its parent's
topics. Each child node is then taken as a parent node in turn, with its left- and right-child nodes 
constructed in the same manner based on splitting the topics in half between the child nodes. 
This procedure is repeated until a binary tree with $K$ leaves (a leaf is a node containing only one topic) 
is obtained, where each node is associated with at least one topic. 

Let the size of each node be the number of latent variables of its associated topics. Hence,
the size of the node associated with topic $k_0$ to topic $k_1$ is $\sum_{k=k_0}^{k_1}n_{dvk}$.
The nested simulation starts from the root of size $C_{dv}$, which contains all topics, and
simulates downwards to obtain sizes for all nodes. The first step samples the size of its left-child (which 
is of size $\sum_{k=1}^{K_{1/2}} n_{dvk}$) and the size of its right-child w.r.t. the 
discrete sampling density
\[
    p(\sum_{k=1}^{K_{1/2}} n_{dvk}=n, \sum_{k=K_{1/2} + 1}^{K} n_{dvk}=C_{dv} - n|\text{rest})=
    \frac{h_{1:K_{1/2}}(n)\times h_{(K_{1/2}+1):K}(C_{dv}-n)}{h_{1:K}(C_{dv})},
\]
for $n=0,1,\ldots,C_{dv}$. Then progressing downwards, sizes for child nodes can be simulated 
given the size of their parent node.
For each parent node associated with topics $k_0$ to $k_1$ ($k_0<k_1$) 
with size $\sum_{k=k_0}^{k_1}n_{dvk} > 0$,
sizes for the children nodes are simulated from the distribution
\begin{multline}
\label{eq:sampler-nested}
    p(\sum_{k=k_0}^{k_{1/2}} n_{dvk}=n, \sum_{k=k_{1/2} + 1}^{k_1} n_{dvk}=c - n | 
    \sum_{k=k_0}^{k_1}n_{dvk}=c,\text{rest})=\\
    \frac{h_{k_0:k_{1/2}}(n)\times h_{(k_{1/2}+1):k_1}(c-n)}{h_{k_0:k_1}(c)},
\end{multline}
for $n=0,1,\ldots,\sum_{k=k_0}^{k_1}n_{dvk}$ where
$k_{1/2} = [(k_0 + k_1)/2]$. Zero size parent nodes can be skipped as no latent
variables belong to topics associated with this node. 
The nested simulation stops when sizes for all leaves are sampled. 
For the leaf node associated with topic $k$, its size determines 
the value of $n_{dvk}$. It is trivial to see that the product of these conditional densities
leads to the target density (\ref{eq:bcgs-table}).

To enable nested simulation, we need to be able to compute the normalising constants 
$h_{k_0:k_1}(c)$ for the root and all parent nodes in the binary tree. 
Due to its factored structure (\ref{eq:normalising-constant}), a upward
summation approach can be used to recursively obtain each value of $h_{k_0:k_1}(c)$.

It is trivial that $h_{k:k}(c)=q_k(c)$ for any $k\leq K$ and $c\leq C_{dv}$. 
The constants $h_{k_0:k_1}$ with $k_0<k_1$ can be sequentially computed through 
the upward recursive equation
\[
    h_{k_0:k_1}(c) = \sum_{n=0}^c h_{k_0:k_{1/2}}(n) \times h_{(k_{1/2}+1):k_1}(c-n).
\]
as each constituent term $h_{k_0:k_{1/2}}(\cdot)$ and $h_{(k_{1/2}+1):k_1}(\cdot)$ will have been 
previously calculated.

\subsection{Computational complexity}

The blocking scheme increases the computational cost of evaluating
the sampling densities, as the full conditional densities of blocked
latent variables (\ref{eq:bcgs-table}) are more complex compared to those 
for single-site updates (\ref{eq:sampler-collapsed}). 
To compute the sampling densities for all $z_{dn}$ in the
block $B_{dv}$ of size $C_{dv}$, the required number of operations
is $O(C_{dv}^{2}K)$, which is quadratic in $C_{dv}$, while
it is $O(C_{dv}K)$ for the single-site collapsed Gibbs
sampler. This quadratic cost is due to the calculation of normalising 
constants (\ref{eq:normalising-constant}), which in fact is computing 
$C_{dv}$-length discrete convolutions.
In practise, $C_{dv}$, the number of appearances of a given word in a document,
is typically not large. When $K\gg C_{dv}$, the extra computational cost
resulting from the blocking scheme will not be significant, as it
is still linear in $K$. 

Given the sampling densities, the simulation cost of the latent variables
is the other contributor to the computational complexity. To update
all $z_{dn}$ in block $B_{dv}$ of size $C_{dv}$, the single-site
sampler requires $C_{dv}$ sequential steps sampling from a $K$-dimensional discrete
distribution, with each step costing $O(K)$ operations.
For the blocked sampler, the backward simulation in theory requires $K-1$
steps, however it can terminate at any step $k\leq K$ if $C_{dv}-n_{dv1}-\cdots n_{dvk}=0$.
In comparison, the nested simulation requires 
at most $O(C_{dv}\log{K})$ steps due to the binary tree structure. Further, in each step, backward simulation
and nested simulation only require $O(C_{dv}-n_{dv1}-\cdots-n_{d,v,k+1}+1)$ 
and $O(n_{dvk_0}+\cdots+n_{dvk_1})$ 
operations respectively to simulate from their conditional density (\ref{eq:sampler-backward} and
\ref{eq:sampler-nested}). These sampling operations have much less complexity than 
sampling from $K$-dimensional discrete distributions when $C_{dv}\ll K$.

In particular when $C_{dv}=1$, i.e. word $v$ only appears
once in document $d$, computational complexity of the single-site sampling scheme and the backward-simulation 
blocked sampling scheme are $O(K)$ , while the nested scheme 
has far less computational complexity. For the computational
cost of evaluating the sampling densities, the required number of
operations for all schemes is $O(K)$ as $1=C_{dv}=C_{dv}^{2}$.
For the simulation cost, the single-site sampler uses one simulation
from a $K$-dimensional discrete distribution. The backward simulation
performs sequential sampling from at most $K-1$ binomial distributions, 
while the nested scheme performs a binary-tree-search style sampling from at most $\log{K}$
binomial distributions.

\section{Experiments}
\label{sec:ex}

The performance of the collapsed Gibbs sampler using the
proposed blocking scheme was evaluated for one simulated and two real 
datasets. Interest is in two aspects of performance:  mixing
efficiency and the time taken to learn the model. 
Blocked collapsed Gibbs samplers (\ref{eq:sampler-backward} and \ref{eq:sampler-nested}) are first compared
to the single-site collapsed Gibbs sampler (\ref{eq:sampler-collapsed}),
and to the data augmentation sampler (\ref{eq:sampler-augment}) using
the simulated dataset in \citet{Griffiths:2004}. All
collapsed samplers are then compared through analyses of two real datasets: the KOS blog entries from
dailykos.com and the NIPS papers dataset from books.nips.cc, both of which are available
for download from the UCI Machine Learning Repository \citep{Lichman:2013}.
We demonstrate that the blocking scheme can on average
achieve substantial improvements in chain mixing over the state of the art single-site 
sampler, with moderate additional computational cost when $K$ is small. As
$K$ becomes larger, the iteration speed of nested-simulation blocked scheme achieves 
and surpasses the performance of the single-site sampler. This indicates that our blocking 
scheme is particularly suitable to models with a large number of topics, $K$.

\subsection{Evaluation method}

We determine the mixing efficiency of a sampler via two metrics, 
evaluated over the realised MCMC sample path. The first one is the logarithm of 
the posterior probability (\ref{eq:posterior-collapsed-word}): 
given the same number of iterations, the
sampler with better mixing efficiency can reach a region with higher
log posterior probability. While the log posterior probability
contains an intractable normalising constant,
we equivalently
evaluate the unnormalised probability
\begin{multline}
\label{eq:level}
\mathrm{log}\,q(z_{1:D}| w_{1:D},\alpha_{1:K},\beta)=\\
    \sum_{d}\sum_{k}\log\Gamma(S_{d\cdot k}+\alpha_{k})+
    \sum_{k}\sum_{v}\log\Gamma(S_{\cdot vk}+\beta)-
    \sum_{k}\log\Gamma(S_{\cdot\cdot k}+V\beta),
\end{multline}
which is equal to $\log\,p(z_{1:D}| w_{1:D},\alpha_{1:K},\beta)$
plus a constant term for any $z_{1:D}$. The unnormalised log posterior
probability is useful for evaluating the speed that the sampler reaches the
region of high posterior probability given some starting
point. The faster it reaches this region, the shorter burnin period
it will have. However, this metric only measures one aspect of mixing
efficiency.

The second mixing efficiency metric is the perplexity \citep{Blei:2003,Wallach:2009b},
which is the probability assigned to unseen data given some training
documents. Let $w_{A}^{\star}=\{w_{d}^{\star};\,d\in A\}$, with 
$w_{d}^{\star}=(w_{d1}^{\star},\ldots,w_{dN_{d}^{\star}}^{\star})$,
represent the collection of unseen words in the corpus in some test set $A$. The perplexity is given by
\begin{equation}
\label{eq:perp}
\text{perp}(w_{A}^{\star}|\theta_{1:D},\phi_{1:K})=
    \exp\left\{ -\frac{\sum_{d\in A}\sum_{v}C_{dv}^{\star}\log(\sum_{k}\theta_{dk}\phi_{kv})}
    {\sum_{d\in A}N_{d}^{\star}}\right\},
\end{equation}
where $C_{dv}^{\star}=\sum_{n}\delta_{v}(w_{dn}^{\star})$ is the
total number of occurrences of the word $v$ in $w_{d}^{\star}$. 
A document completion approach \citep{Wallach:2009b} is used to partition
each document in the test set $A$ into two sets of words, $w_{d}$ and
$w_{d}^{\star}$, using $w_{d}$ to estimate $\theta_{d}$, and then
calculating the perplexity on $w_{d}^{\star}$. In the $l^{\text{th}}$
MCMC iteration, $\theta_{dk}$ and $\phi_{kv}$ are respectively estimated by 
\[
\theta_{dk}^{(l)}=\frac{S_{d\cdot k}^{(l)}+\alpha_{k}}{N_{d}+\sum_{k}\alpha_{k}}\quad\text{ and }\quad
\phi_{kv}^{(l)}=\frac{S_{\cdot vk}^{(l)}+\beta}{S_{\cdot\cdot k}^{(l)}+V\beta}.
\]
For every $L=10$ iterations, the sample mean
$\frac{1}{L}\sum_{l}\left(\sum_{k}\theta_{dk}^{(l)}\phi_{kv}^{(l)}\right)$
is used to estimated the $\sum_{k}\theta_{dk}\phi_{kv}$ term in (\ref{eq:perp}),
in order to circumvent the label switching problem \citep{Celeux:2000} in
MCMC samples. The perplexity can measure the ability of a sampler
to explore the posterior density.
The sampler which better explores this region can provide estimators
with better predictive performance, and thereby a lower perplexity
value.

\begin{figure}
\centering
\includegraphics[scale=0.5]{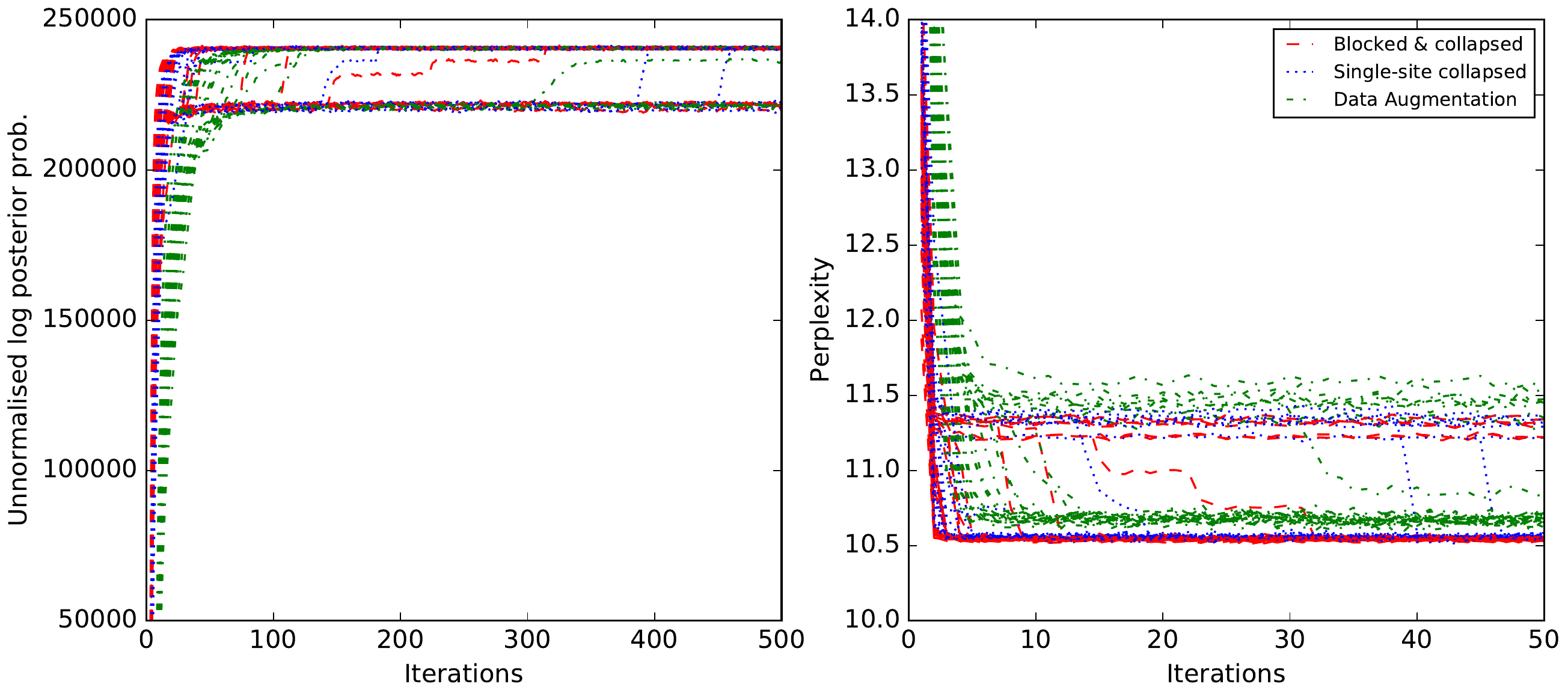}
\caption{\label{fig:synth}\small The unnormalised log posterior probability (left)
and the perplexity (right) of the blocked collapsed Gibbs sampler 
(dashed), the single-site collapsed Gibbs sampler (dotted), and the data augmentation 
sampler (dashed \& dotted) for the simulated dataset ($D=2,000,\,V=25,\,K=10$).}
\end{figure}

\subsection{Results for simulated data}

We analyse a simulated corpus of $D=2,000$ documents following the
model setup of \citet{Griffiths:2004}, with $K=10$ topics for all
documents and $V=25$ unique words in the vocabulary. Each document
in the simulated corpus has $100$ words, and the hyper-parameters
in the model are set to be $\alpha_{k}=0.1$ for all $k$ and $\beta=0.01$.
To evaluate the perplexity, we hold out half of the words in the last
$250$ documents as the test dataset, with the remainder as training
data. That is, $N_{d}=100$ for $d\leq1,750$ and $N_{d}=N_{d}^{\star}=50$
for $d>1,750$. We implement the blocked and single-site collapsed
Gibbs samplers and the data augmentation sampler 30 times, identically initialised
at random points. The estimates of the unnormalised log posterior
probability (\ref{eq:level}) and the perplexity (\ref{eq:perp})
are shown in Figure~\ref{fig:synth}.

For this small dataset, while most runs of each sampler appear to converge within
$\sim 100$ iterations, not all converge to the region of the global
posterior mode. Both collapsed samplers perform well on average, with
relatively few runs (5 out of 30) trapped in regions of local modes after 500 iterations, 
compared to the data augmentation sampler (10 out of 30 runs).
Both collapsed
samplers consistently outperform the data augmentation algorithm in
achieving lower perplexity. 
As the burnin efficiency of the two collapsed samplers 
is very rapid for most runs
($\sim50$ iterations), the blocked sampler only achieves slightly better performance for this dataset.

While this small dataset does not demonstrate a clear
advantage for the blocked collapsed Gibbs sampler over the single-site
sampler, it does illustrate that sampler performance can improve even in non-challenging scenarios.
However in practice,
real datasets are commonly both large and sparse, such that the single-site
collapsed Gibbs sampler performs poorly. 

\begin{figure}
\centering
\subfloat[\label{fig:kos}KOS]{\includegraphics[scale=0.525]{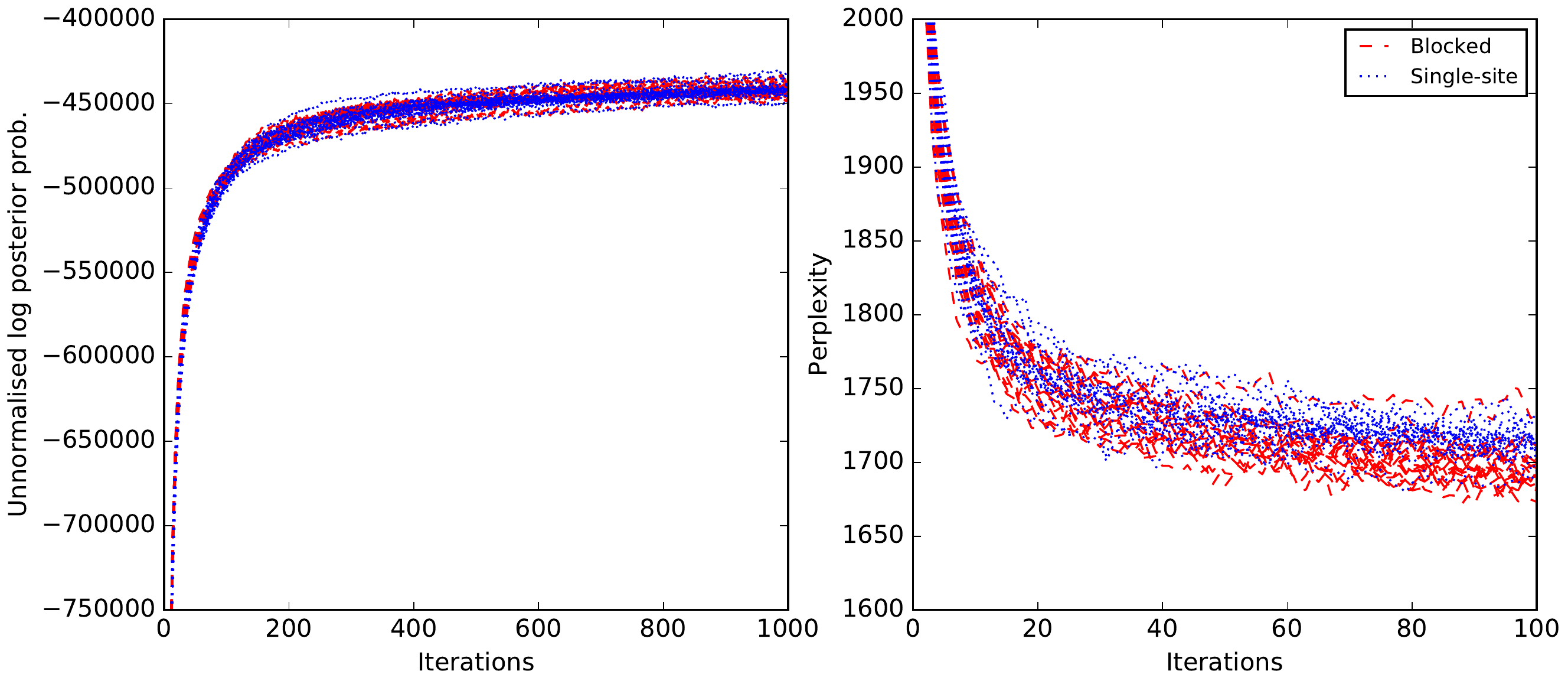}}\\
\subfloat[\label{fig:nips}NIPS]{\includegraphics[scale=0.525]{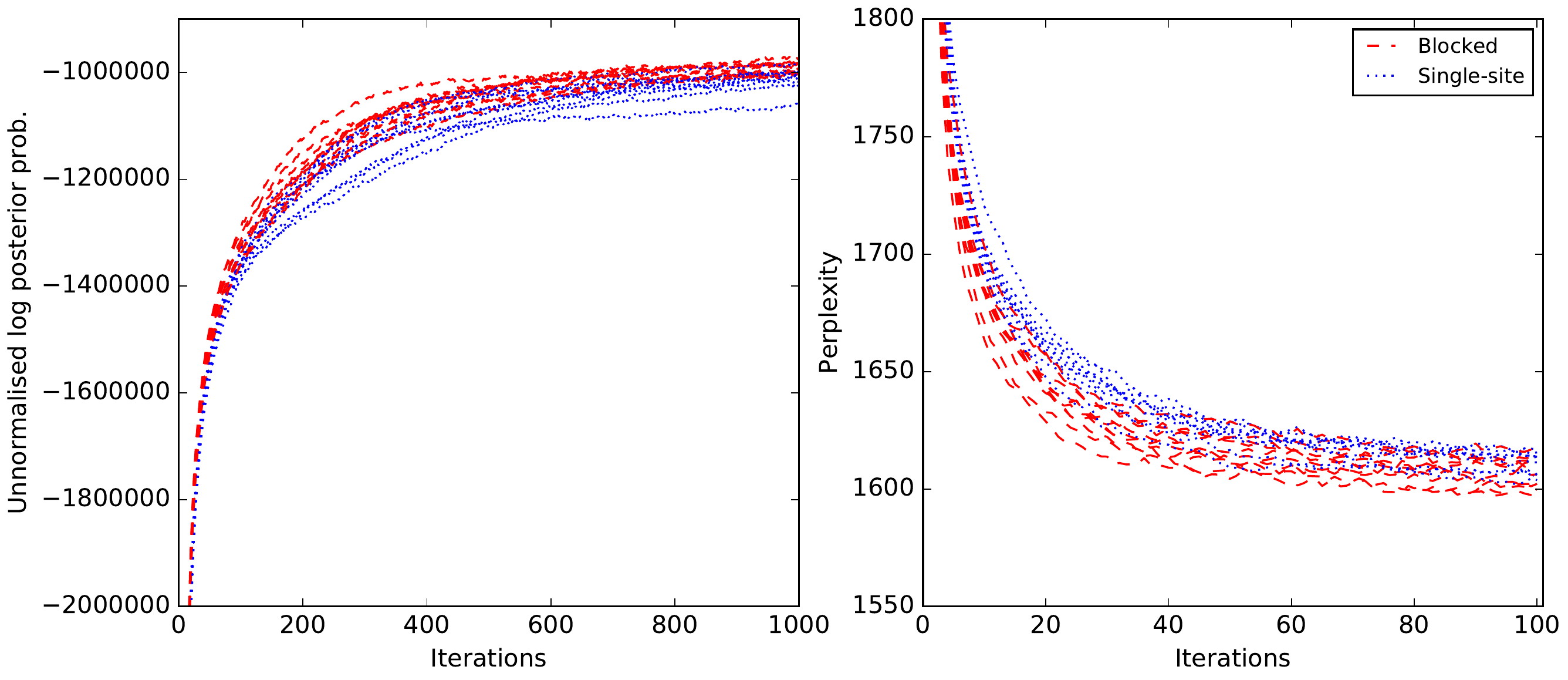}}
\caption{\small The unnormalised log posterior probability (left) and the perplexity
(right) of the collapsed Gibbs samplers with blocking (dashed) and
singe-site update (dotted) schemes for the KOS dataset with $K=32$
(top), and NIPS dataset with $K=40$ (bottom).}
\end{figure}

\subsection{Results for real data}

We analyse two corpora, the KOS corpus of $D=3,430$ documents with
$V=6,906$ unique words and $K=32$ topics, and the NIPS corpus of
$D=1,500$, $V=12,419$ and $K=40$. The KOS (NIPS) corpus has $467,714$ ($1,932,365$)
total words. As before, the LDA model hyper-parameters are set to be $\alpha_{k}=0.1$
for all $k$ and $\beta=0.01$. To evaluate the perplexity, we hold
out half of the words in the last 430 (250) documents of the KOS(NIPS)
dataset as test data, with the remainder used as training data. We
implement the blocked and single-site collapsed Gibbs samplers 20 (10)
times, initialised at random points, for the KOS (NIPS) dataset. The
estimates of the unnormalised log posterior probability (\ref{eq:level})
and the perplexity (\ref{eq:perp}) are shown in Figure~\ref{fig:kos}
(KOS) and Figure~\ref{fig:nips} (NIPS). 

For the KOS dataset, the blocked sampler performs no better than the
single-site sampler in terms of unnormalised log posterior probability,
while it performs a little better in terms of  perplexity. This is expected to occur
as $62.9\%$ of words in the KOS dataset appear only once in their
documents. As a result, most blocks will have only one latent variable,
and so sampling such blocks is equivalent to sampling a single variable,
as for the single-site collapsed Gibbs sampler. Therefore, the gain
in performance achieved by blocking, while apparent, is not particularly large.

For the NIPS dataset, the performance of the blocked sampler is significantly
better than the single-site sampler. The blocked sampler can reach the region
of high posterior probability several hundred iterations faster than the single site sampler.
Further, the blocked sampler achieves lower perplexity values on average, due to the 
blocking scheme which enables a more efficient exploration of the posterior density.

\subsection{Time comparison results}

\begin{figure}
\centering
\subfloat[\label{time-kos}KOS]{\includegraphics[scale=0.4]{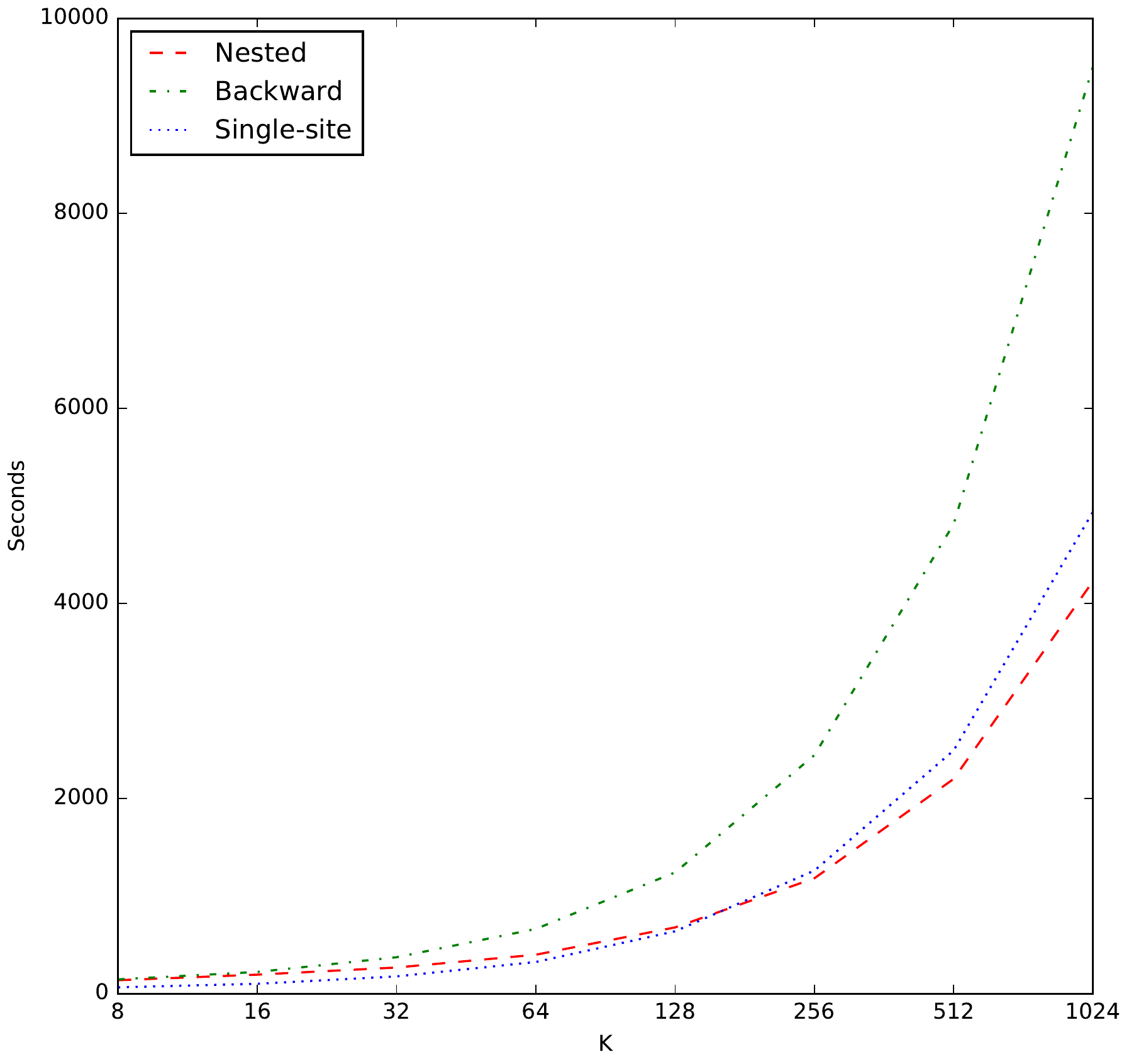}}
\hfill{}\subfloat[\label{time-nips}NIPS]{\includegraphics[scale=0.4]{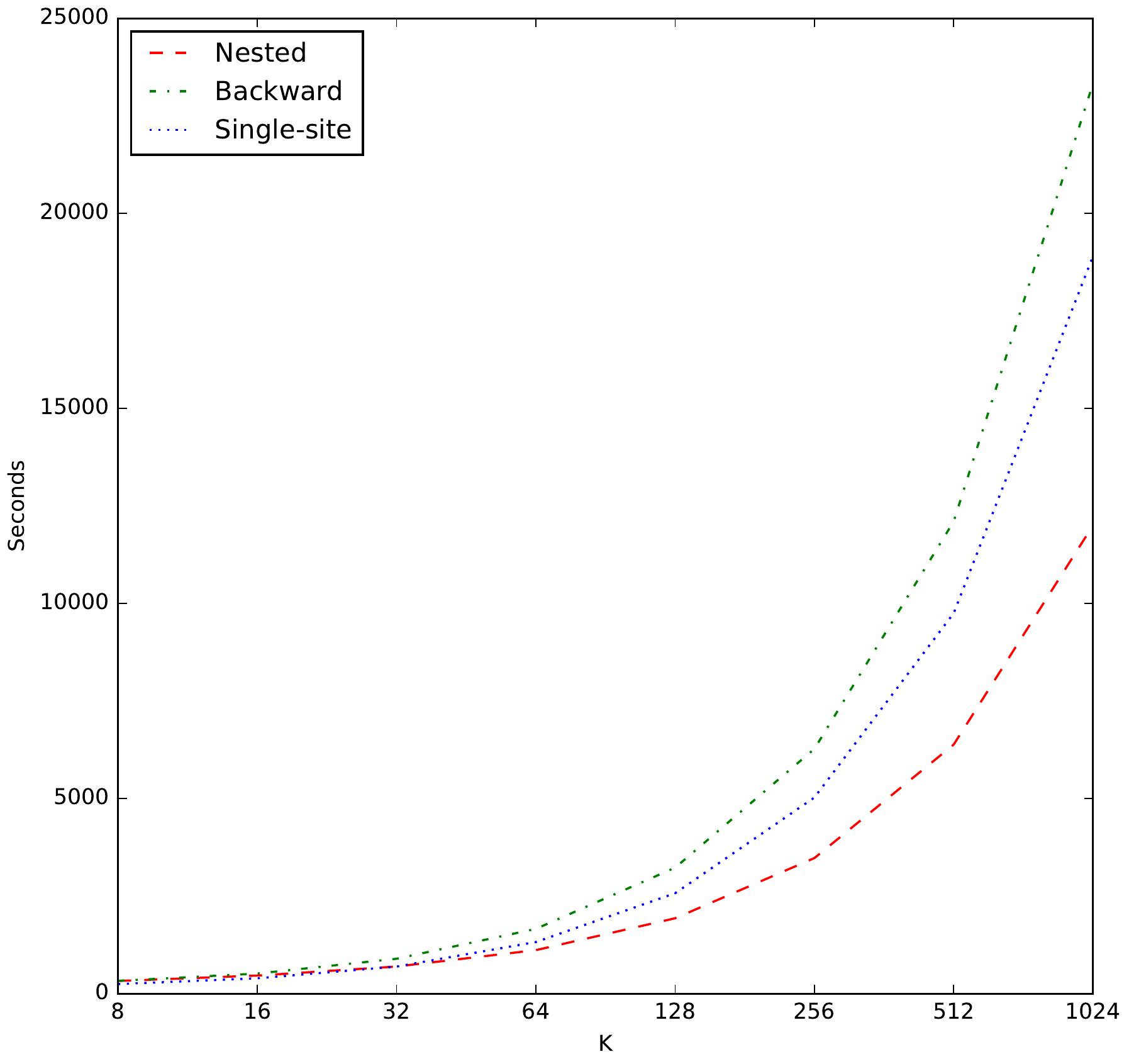}}
\caption{\small Average time per iteration (in seconds) for 
Gibbs samplers with nested blocking (dashed line), backward blocking (dashed \& dotted line) and 
single-site update (dotted line)
schemes, for KOS (left) and NIPS (right) datasets with $K$ from 8 to 1024. \label{fig:time}}
\end{figure}

To investigate the computational costs of the different collapsed Gibbs samplers,
we implement  blocking (nested- and backward-simulation) and single-site schemes
algorithms (in Python 3.4) on the above two real datasets, replicated 10 times
under different settings on a cluster node with one CPU core with
the Intel Xeon E5-2670 2.60 GHz processor and 12 Gb RAM.
The average time costs per iteration, measured in seconds, for each
algorithm on the KOS and NIPS datasets for models with $K=8,16,\ldots,1024$ topics
are shown in Figure~\ref{fig:time}.

For the smaller KOS dataset, the single-site collapsed
sampler runs at more than twice the speed of the backward-simulation blocked sampler for any number of topics. For
the relatively larger NIPS dataset, the backward-simulation method takes around
25\% extra computation time. However, the nested-simulation blocked sampler, though
only slightly better than the backward-simulation scheme when $K$ is around 16, 
achieves and surpasses the performance of the single-site sampler as $K$ becomes larger.
In particular when $K=1024$, the nested-simulation blocked sampler can save 14.4\% and 
36.5\% computation per iteration over the single-site sampler for the KOS and NIPS datasets, respectively.
This indicates that our blocking sampler can achieve better both mixing with lower computational cost.

\section{Discussion}
\label{sec:disc}

We have introduced a novel blocking scheme for the collapsed
Gibbs sampler applied to the LDA model, which can, with a theoretical guarantee,
improve chain mixing \citep{Liu:1994a}. Our approach uses a backward simulation or nested simulation scheme
to directly sample from the conditional distributions of blocked latent variables.
We have demonstrated that the blocked collapsed sampler can achieve substantial improvements in
chain mixing, compared to the state of the art single-site collapsed
Gibbs sampler, with the nested-simulation method taking significant less computational cost
for models with more than hundreds of topics.

Various directions could be explored to further reduce the computation cost for sampling
each block. A more efficient simulation procedure could take topic sparsity and
algorithm parallelisation into account. In addition, the $O(C^2_{dv})$  quadratic cost for evaluating the
sampling densities can be to reduced to $O(C_{dv} \log(C_{dv}))$ by using
a fast Fourier transformation based discrete convolution when $C_{dv}$ is large. 
It may not be possible to reduce this further to a linear cost without 
making an approximation. An $O(C_{dv})$ approximation to sample the block without sacrificing 
much efficiency is worth investigation, however.

Another research direction is to turn the proposed blocking scheme
into a general methodology and extend it to other models with mixture
structures. One specific possibility under investigation is to design 
blocking schemes for the marginal sampler of Dirichlet process mixture models 
\citep{Neal:2000} and  hierarchical Dirichlet process models \citep{Teh:2006a}
with discrete observations and conjugate priors.

\section*{Acknowledgements}

\noindent We thank the reviewers for valuable comments. 
This research was supported by the Australian Research Council (DP160102544). 
Xin Zhang was supported by the China Scholarship Council.

\bibliographystyle{chicago}
\bibliography{blocklda}

\end{document}